\documentclass[a4paper,nofootinbib,pdftex,pra,showkeys,showpacs,superscriptaddress,twocolumn]{revtex4}

\pdfoutput=1

\usepackage{dcolumn}
\usepackage{graphicx}
\usepackage{pxfonts}
\usepackage{textcomp}
\usepackage[hyperfootnotes=false]{hyperref}
\usepackage{hypernat}

\begin{document}

\title{Precise dipole moments and quadrupole coupling constants of the cis and trans conformers of
   3-aminophenol: Determination of the absolute conformation}%
\author{Frank Filsinger}%
\author{Kirstin Wohlfart}%
\author{Melanie Schnell}%
\affiliation{Fritz-Haber-Institut der Max-Planck-Gesellschaft, Faradayweg 4--6, 14195 Berlin,
   Germany}%
\author{Jens-Uwe Grabow}%
\affiliation{Gottfried-Wilhelm-Leibniz-Universit\"at, Institut f\"ur Physikalische Chemie und
   Elektrochemie, Callinstra{\ss}e 3a, 30167 Hannover, Germany}%
\author{Jochen K\"upper}%
\email[Author to whom correspondence should be addressed. Electronic mail:
]{jochen@fhi-berlin.mpg.de}%
\affiliation{Fritz-Haber-Institut der Max-Planck-Gesellschaft, Faradayweg 4--6, 14195 Berlin,
   Germany}%
\date{\today}%
\pacs{33.20.Bx; 33.55.Be; 33.15.-e}%
\keywords{microwave spectroscopy; dipole moment; Stark effect; conformers; cold molecules;
   rotational spectroscopy; supersonic jet; molecular structure}%
\begin{abstract}
   The rotational constants and the nitrogen nuclear quadrupole coupling constants of
   cis-3-aminophenol and trans-3-aminophenol are determined using Fourier-transform microwave
   spectroscopy. We examine several $J=2\leftarrow{}1$ and $1\leftarrow{}0$ hyperfine-resolved
   rotational transitions for both conformers. The transitions are fit to a rigid rotor Hamiltonian
   including nuclear quadrupole coupling to account for the $^{14}$N nuclear spin. For cis-3-aminophenol
   we obtain rotational constants of $A=3734.930$~MHz, $B=1823.2095$~MHz, and $C=1226.493$~MHz, for
   trans-3-aminophenol of $A=3730.1676$~MHz, $B=1828.25774$~MHz, and $C=1228.1948$~MHz. The dipole
   moments are precisely determined using Stark effect measurements for several hyperfine
   transitions to $\mu_a=1.7718$~D, $\mu_b=1.517$~D for cis-3-aminophenol and $\mu_a=0.5563$~D,
   $\mu_b=0.5375$~D for trans-3-aminophenol. Whereas the rotational constants and quadrupole
   coupling constants do not allow to determinate the absolute configuration of the two conformers,
   this assignment is straight-forward based on the dipole moments. High-level \emph{ab initio}
   calculations (B3LYP/6-31G$^*$ to MP2/aug-cc-pVTZ) are performed providing error estimates of
   rotational constants and dipole moments obtained for large molecules by these theoretical
   methods.
\end{abstract}
\maketitle%

\section{Introduction}
\label{sec:introduction}

Since the observation of multiple conformers of tryptophan in a supersonic jet at low temperatures
20~years ago~\cite{Rizzo:JCP83:4819}, such occurrence of multiple conformers (torsional isomers,
rotamers) is quite common for the so called ``building blocks of life under isolated
conditions''~\cite{Weinkauf:EPJD20:309, Simons:PCCP6:E7} and other modular molecules, even at the
low temperatures in a cold supersonic jet. Vast progress was made on the electronic and vibrational
spectroscopy of these species. Often the comparison of experimental and \emph{ab initio} vibrational
frequencies can be used to distinguish between the isomers~\cite{Weinkauf:EPJD20:309,
   Simons:PCCP6:E7}. However, in many cases the structural differences are subtle, resulting in very
similar vibrational spectra. Thus more sophisticated methods are required. Dong and Miller, for
example, assigned individual isomers of cytosine exploiting the angles between vibrational
transition moments and the permanent dipole moments of oriented
cytosine~\cite{Dong:Science298:1227}.

Rotational spectroscopy provides precise moments of inertia (rotational constants), quadrupole
coupling constants, and dipole moments, which allow for a detailed understanding of the structural
and electronic properties of individual conformers. In principle, all of these parameters can be
compared to results of \emph{ab initio} calculations in order to assign the observed species to
calculated structures. However, for structurally similar isomers (\emph{vide supra}), i.\,e., when
only the positions of some hydrogen atoms are different between the isomers, also the rotational
constants cannot be used to assign the individual isomer. In such cases, the nuclear quadrupole
coupling constants determined from FTMW spectroscopy have been used to discriminate between
different calculated minimum structures of amino acids~\cite{Lesarri:ACIE43:605}. Similarly, the
permanent dipole moments of the isomers can be used to unambiguously assign individual
conformers~\cite{Reese:JACS126:11387}. Moreover, in principle isotopic substitution combined with a
Kraitchman analysis~\cite{Gordy:MWMolSpec} can be applied to derive conformational information. One
has to carefully consider, however, that the substitution can also induce structural changes
itself~\cite{Giuliano:AC117:609}.

\begin{figure}
   \centering
   \includegraphics[width=\linewidth]{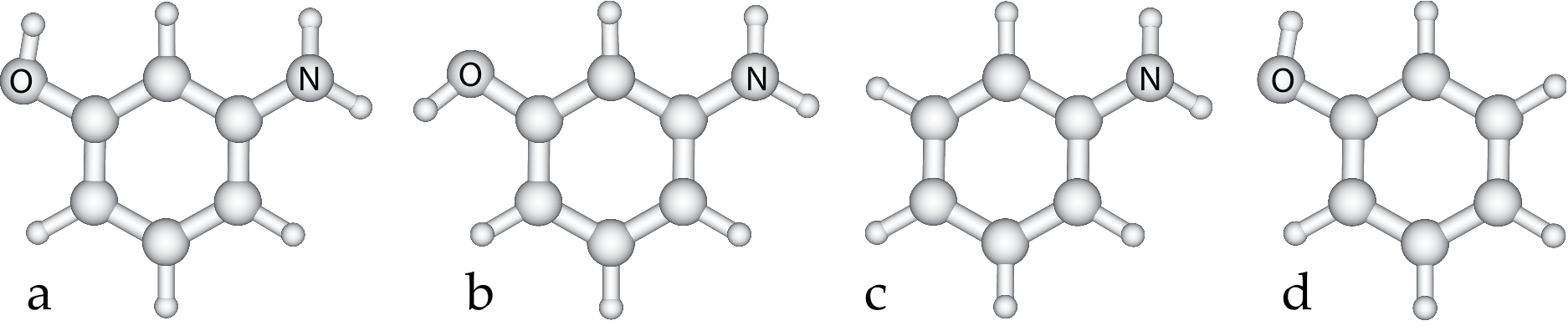}
   \caption{Structures of (a) cis- and (b) trans-3-aminophenol, compared to (c) aniline and (d)
      phenol}
   \label{fig:structures}
\end{figure}

Here we present a detailed study of the individual conformers of 3-aminophenol (3AP) using
Fourier-transform microwave spectroscopy (FTMW). 3AP exists in two distinct conformational
configurations, cis-3-aminophenol (c3AP) and trans-3-aminophenol (t3AP), as shown in
Figure~\ref{fig:structures}.
3AP is a very interesting model system for studying primary molecular properties, because the two
conformers are structurally very similar, varying only in the position of one single hydrogen atom,
but the electronic properties are quite distinct. The two conformers have very
similar rotational constants, but quite different dipole moments and, thus, Stark
effects~\cite{Reese:JACS126:11387}. The individual conformers can also easily be selectively
detected by REMPI-spectroscopy due to their different electronic
properties~\cite{Unterberg:CP304:237, Shinozaki:PCCP5:5044}. From a comparison of experimental
vibrational frequencies and high-level \emph{ab initio} calculations an assignment of the
conformations was possible~\cite{Unterberg:CP304:237}. Moreover, 3AP is the chromophore of the
essential amino acid tyrosine and closely related to dopamine, which also is a benzene derivative
with a phenol group (OH) and an amino group (NH$_2$) substituent. From a detailed understanding of
3AP one can proceed to studies of these more complicated molecules.

The rotational spectra of the parent molecules phenol and aniline, shown in
Figure~\ref{fig:structures}, have been extensively studied using microwave spectroscopy, including
the hyperfine structure due to the nitrogen nuclear quadrupole moment for aniline
\cite{Kleiboemer:ZNatA43:561} and the full substitution structures \cite{Lister:JMolStruct23:253,
   Larsen:JMolStruct51:175}. The dipole moments of both species were also determined using Stark
effect measurements \cite{Larsen:JMolStruct51:175, Lister:JMolStruct23:253}. For 3-aminophenol,
however, to our knowledge no microwave spectroscopy investigation has been performed. The rotationally resolved
electronic excitation spectrum was obtained using high-resolution laser-induced fluorescence
spectroscopy and the dipole moment was determined from Stark effect measurements of these spectra
\cite{Reese:JACS126:11387}, but this study did not provide any details on the hyperfine structure
due to the nitrogen nuclear quadrupole moment.

In order to obtain the rotational constants and the dipole moment with high accuracy, we determine
the rotational constants, nuclear quadrupole coupling constants, and dipole moment components of
c3AP and t3AP using high-resolution FTMW spectroscopy without and with applied electric fields in
the coaxially oriented beam-resonator arrangement (COBRA). We perform these measurements for the
very lowest rotational states, because these states have the most sensitive Stark
effect~\cite{Gordy:MWMolSpec}. In the Stark effect measurements we give special attention to an
accurate calibration of the electric field strengths, which is detailed in
Appendix~\ref{sec:calibration}.

In addition, we perform high-level \emph{ab initio} calculations to test the quality of theoretical
descriptions for these molecules. The comparisons of theoretical and experimental results will allow
us to estimate the errors of theoretical rotational constants and dipole moments of similar
molecules we might study in the future.

\section{Experimental details}
\label{sec:experimental}

The experimental setup of the Hannover COBRA-FTMW-spectrometer is described in detail
elsewhere~\cite{Grabow:RSI67:4072, Schnell:RSI75:2111}. In brief, 3-aminophenol (purity
$\geq98$\,\%) was purchased from Sigma-Aldrich and used without further purification. The sample was
heated to 120\,\textdegree{C} and co-expanded in 2~bar of Ne through a pulsed nozzle (General Valve
Series 9) with a 0.8~mm orifice. The supersonic expansion was pulsed coaxially into the microwave
resonator~\cite{Grabow:RSI67:4072}, which was specially developed to provide high sensitivity and
resolution at low frequencies down to 2~GHz, and the lowest rotational transitions of 3AP in the
range of 3--7.5~GHz were recorded with a linewidth (FWHM) of 2.5~kHz and a frequency accuracy of
500~Hz.

\begin{figure}
   \centering
   \includegraphics[width=\linewidth]{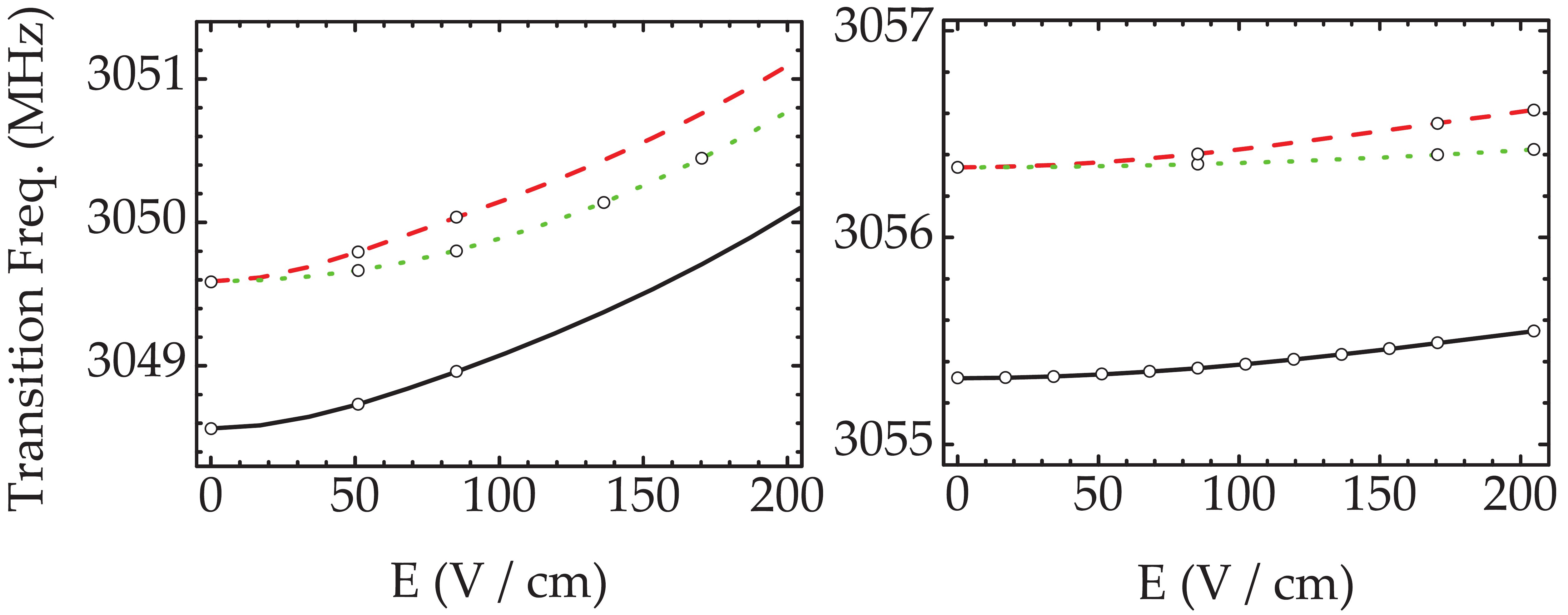} 
   \caption{(Color online:) Stark shift measurements (circles) and fitted Stark-lobes (lines) for
      cis-3-aminophenol (left) and trans-3-aminophenol (right). For both molecules the
      $J_{K_aK_c}=1_{01}\leftarrow{}0_{00}, F=0\leftarrow{}1, m_F=0\leftarrow{}1$ (solid black
      lines), $F=2\leftarrow{}1, m_F=1\leftarrow{}0$ (dashed red lines), and $F=2\leftarrow{}1,
      m_F=2\leftarrow{}1$ (dotted green lines) were measured.}
   \label{fig:starkshift}
\end{figure}

Stark shift measurements were performed with the Coaxially Aligned Electrodes for Stark-effect
Applied in Resonators (CAESAR) setup~\cite{Schnell:RSI75:2111}. This setup provides a homogeneous
electric field over the entire volume, from which molecules are effectively contributing to the
emission signal. We calibrated the field strength using the $J=1\leftarrow{}0$ transition of
OC$^{36}$S (0.02\,\% natural abundance) using a documented dipole moment of
0.71519\,(3)~D~\cite{Reinartz:CPL24:346}, see Appendix~\ref{sec:calibration} for details. In this
way the dipole moment components were determined from several hyperfine transitions of the
$J=1\leftarrow0$ band measured at different electric field strengths up to 205~V/cm. The individual
measurements and the respective fitted Stark lobes are shown in Figure \ref{fig:starkshift}.

\section{Computational details}
\label{sec:simulations}
% other \emph{ab initio studies}: \cite{Brinck:JACS119_4239, Gomes:IJQC101:860,
%    Robinson:JPCA108:4420, Unterberg:CP304:237}

We performed \emph{ab initio} calculations of c3AP and t3AP using the Gaussian~03 program package
\cite{Gaussian:2003C02}. Previously, rotational constants and dipole moments had been calculated
using the B3PW91/6-31G$^*$ \cite{Reese:JACS126:11387} and CASSCF and CASPT2/6-31G$^*$
\cite{Unterberg:CP304:237} levels of theory. We extended these calculations to multiple methods
(B3LYP, B3PW91, MP2) using the same 6-31G$^*$ and the larger aug-cc-pVTZ basis sets. We performed
fully relaxed geometry optimizations of both conformers at these levels of theory and calculated the
dipole moments for the obtained minimum structures on the potential energy surface. However, this
procedure yielded considerable out-of-plane $\mu_c$ dipole moment components due to the non-planar
structure around the nitrogen nucleus and the corresponding out-of-plane non-bonding $sp^3$ orbital
on the nitrogen atom. These $\mu_c$ dipole moment components are, however, not experimentally
observable as the zero-point vibrational level averages over all out-of-plane angles due to the
inversion tunneling of the two amino hydrogen atoms. The tunneling rate is similar to the one of
aniline (1~THz~\cite{Kleiboemer:ZNatA43:561}) and much larger than the rotational frequency. To test
the calculated dipole moments we also performed geometry optimizations for the planar transition
states of the inversion motion for both conformers in $C_s$ symmetry. These calculations confirmed
the previous reasoning and gave in-plane dipole moment components quite similar to the specified
ones. The obtained rotational constants and dipole moments are presented together with the
experimental results in Section~\ref{sec:results}. For an improved theoretical description, one
would need to calculate the dipole moment function along the inversion motion coordinate and derive
the vibrationally averaged expectation values $\left<\mu_\alpha\right>$, which is beyond the scope
of this work.

\section{Results and discussion}
\label{sec:results}

\begin{figure}
   \centering
   \includegraphics[width=\linewidth]{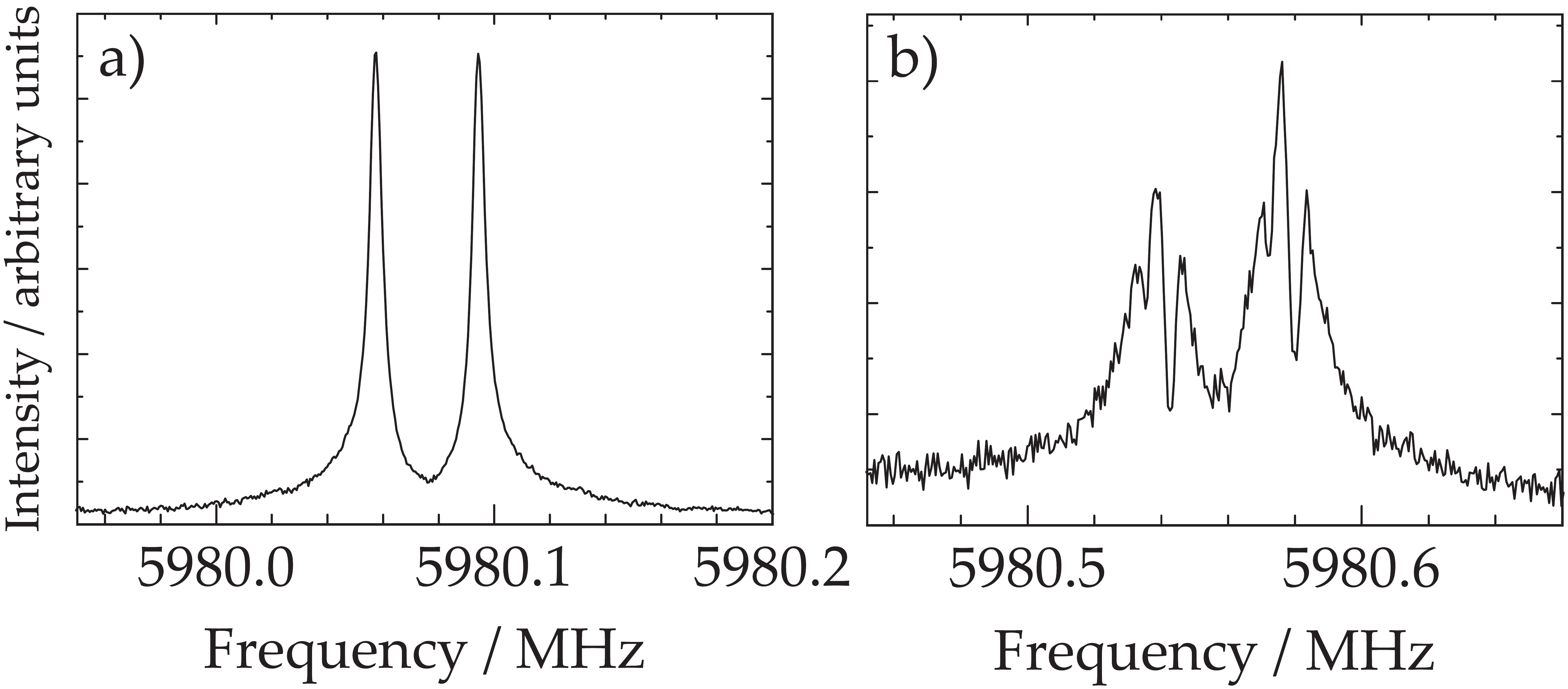}
   \caption{Amplitude spectrum of the
      $J_{K_aK_c}=2_{02}3\leftarrow1_{01}2$ (a) and
      $2_{02}1\leftarrow1_{01}0$ (b) hyperfine transitions for c3AP. The major doublet-splitting in
      both Figures is due to the Doppler-splitting in the coaxial spectrometer-arrangement.
      Additional splittings of the individual Doppler-components of this single quadrupole-component
      of the rotational transition are obvious in Figure (b). We attribute these splittings to
      partially resolved magnetic hyperfine structure due to the hydrogen nuclear spins; see text
      for details.}
   \label{fig:splitting}
\end{figure}

The measured field-free microwave transitions of c3AP and t3AP are given in
Tables~\ref{tab:lines:cis} and \ref{tab:lines:trans}, respectively.
\begin{table}
   \centering
   \begin{tabular}{cccc@{ $\leftarrow$ }ccdd}
      \hline\hline
      $J'_{K_a'K_c'}$ &{ $\leftarrow$ }& $J''_{K_a''K_c''}$ & $F'$ & $F''$ & comp.
      & \multicolumn{1}{c}{obs.~(MHz)} & \multicolumn{1}{c}{obs.$-$calc.~(MHz)} \\
      \hline
      $1_{01}$ &{ $\leftarrow$ }  &$0_{00}$ & 0 & 1 & 1 & 3048.5622 & -0.0017 \\
               &                  &         & 2 & 1 & 1 & 3049.5857 & -0.0030 \\
               &                  &         & 1 & 1 & 1 & 3050.2767 &  0.0047 \\
      $2_{02}$ &{ $\leftarrow$ }  &$1_{01}$ & 1 & 1 & 3 & 5978.8532 &  0.0009 \\
               &                  &         & 3 & 2 & 1 & 5980.0761 & -0.0005 \\
               &                  &         & 2 & 1 & 1 & 5980.3687 &  0.0014 \\
               &                  &         & 1 & 0 & 3 & 5980.5581 & -0.0023 \\
               &                  &         & 2 & 2 & 2 & 5981.0511 &  0.0006 \\       
      $2_{12}$ &{ $\leftarrow$ }  &$1_{01}$ & 1 & 1 & 2 & 7412.7858 &  0.0020 \\
               &                  &         & 3 & 2 & 3 & 7414.2197 & -0.0024 \\ 
               &                  &         & 2 & 1 & 2 & 7414.8910 &  0.0021 \\ 
               &                  &         & 2 & 2 & 1 & 7415.5794 & -0.0017 \\ 
      \hline\hline
   \end{tabular}
   \caption{Measured hyperfine-split transitions for cis-3-aminophenol with fit residuals. Also the
      number of clearly split components is given for each transition. Note that additional
      splittings, which are not resolved by our spectrometer, might be present. See text for details. }
   \label{tab:lines:cis}
\end{table}
\begin{table}
   \centering
   \begin{tabular}{cccc@{ $\leftarrow$ }ccdd}
      \hline\hline
      $J'_{K_a'K_c'}$ &{ $\leftarrow$ }& $J''_{K_a''K_c''}$ & $F'$ & $F''$ & comp.
      & \multicolumn{1}{c}{obs.~(MHz)} & \multicolumn{1}{c}{obs.$-$calc.~(MHz)} \\
      \hline
      $1_{01}$ &{ $\leftarrow$ }  &$0_{00}$ & 0 & 1 & 1 & 3055.3215 &  0.0021 \\
               &                  &         & 2 & 1 & 1 & 3056.3392 & -0.0001 \\
               &                  &         & 1 & 1 & 2 & 3057.0218 &  0.0025 \\
      $1_{11}$ &{ $\leftarrow$ }  &$0_{00}$ & 0 & 1 & 1 & 4957.3704 &  0.0011 \\
               &                  &         & 2 & 1 & 2 & 4958.2638 &  0.0006 \\
      $2_{02}$ &{ $\leftarrow$ }  &$1_{01}$ & 1 & 1 & 1 & 5990.5985 &  0.0011 \\
               &                  &         & 3 & 2 & 1 & 5991.8183 & -0.0007 \\
               &                  &         & 2 & 1 & 2 & 5992.1126 & -0.0014 \\
               &                  &         & 1 & 0 & 1 & 5992.2975 &  0.0003 \\
               &                  &         & 2 & 2 & 2 & 5992.7950 &  0.0010 \\       
      $2_{11}$ &{ $\leftarrow$ }  &$1_{10}$ & 3 & 2 & 1 & 6712.8958 & -0.0016 \\
      $2_{12}$ &{ $\leftarrow$ }  &$1_{01}$ & 1 & 1 & 2 & 7413.1225 &  0.0001 \\
               &                  &         & 1 & 2 & 1 & 7413.8012 & -0.0011 \\
               &                  &         & 1 & 0 & 1 & 7414.8210 & -0.0013 \\ 
               &                  &         & 2 & 1 & 3 & 7415.2475 & -0.0016 \\ 
               &                  &         & 2 & 2 & 1 & 7415.9312 &  0.0022 \\
      \hline\hline
   \end{tabular}
   \caption{Measured hyperfine-split transitions for trans-3-aminophenol with fit residuals. Also the
      number of clearly split components is given for each transition. Note that additional
      splittings, which are not resolved by our spectrometer, might be present. See text for details.}
   \label{tab:lines:trans}
\end{table}
For each conformer, all hyperfine transitions were fit to a rigid-rotor Hamiltonian, including
nuclear quadrupole coupling for the nitrogen nucleus, using the computer program
QStark~\cite{Kisiel:CPL325:523,Kisiel:JPC104:6970}. Several lines show additional splittings on the
order of 5--10~kHz, as depicted in Figure~\ref{fig:splitting}~b.
The number of clearly visible split components is given in Tables~\ref{tab:lines:cis} and
\ref{tab:lines:trans}. Many lines show additional shoulders which are not included in the Tables, as
it is not possible to reliably determine them. These splittings are most likely due to magnetic
spin-spin or spin-rotation coupling of the hydrogen atoms, which are not included in the used
Hamiltonian. The magnetic-interaction energy of two hydrogen nuclei with parallel spins is
approximately 6~kHz for a distance of 170~pm,\footnote{The interaction energy of two magnetic
   dipoles is given by $W=\frac{\mu_0\cdot\mu\cdot\mu'}{4\pi\cdot{r^3}}$.} which corresponds to the
distance of the two amine hydrogen atoms. The distance between the hydroxy hydrogen to the closest
hydrogen atom on the ring is 230~pm yielding a somewhat smaller interaction energy of approximately
2.5~kHz. The spin-rotation interaction of the two amine hydrogen atoms, for example, can be
estimated based on the NH$_3$ coupling constant (18.5~kHz~\cite{Kukolich:PR156:83}), scaled by the
relative angular velocity of ammonia ($B+C\approx\text{500~GHz}$) and 3AP
($B+C\approx\text{3~GHz}$). The approximately 180 times smaller angular velocity of 3AP compared to
ammonia suggests couplings, which are also two orders of magnitude smaller. On the other hand, the
wider charge distribution of 3AP increases the induced magnetic field and, therefore, the coupling
strength. In summary, we estimate a spin-rotation coupling on the order of 1~kHz. From the number of
resolved components for the different $J'\leftarrow{}J''$, given in Tables~\ref{tab:lines:cis} and
\ref{tab:lines:trans}, no systematic dependence on $J$ can be derived. Splittings due to
spin-rotation coupling are expected to increase with $J$, while for spin-spin coupling, one would
expect a slight decrease of the splitting with increasing $J$~\cite{Gordy:MWMolSpec}. Both effects
are expected to result in a splitting comparable to the observed one. A detailed evaluation of these
splittings would require even higher resolution measurements, the examination of higher-$J$
transitions, or, most probably, both, in order to sufficiently resolve and analyze the underlying
hyperfine structure, as has been done for ammonia \cite{Kukolich:PR138:1322, Kukolich:PR156:83,
   Veldhoven:EPJD31:337}. For the purpose of this work, we determined the line-center frequencies as
the intensity-weighted average of these split lines.

The fit to the experimental data is very good with remaining standard deviations of 2.93~kHz and
1.65~kHz for c3AP and t3AP, respectively. This is somewhat higher than usual, because the
experimental frequencies were obtained from intensity weighting of the underlying unassigned
additional hyperfine structure (\emph{vide supra}). We want to point out that for c3AP twelve
hyperfine-split lines from only three rotational transition are evaluated to yield three rotational
constants and two quadrupole coupling constants. Using the
QStark~~\cite{Kisiel:CPL325:523,Kisiel:JPC104:6970} program we perform a global fit by direct
diagonalization of the Hamiltonian matrix including terms for all
parameters~\cite{Albritton:LeastSquaresFitting}. This determines well-defined values and error
estimates for all five parameter simultaneously. The obtained rotational constants and quadrupole
coupling constants are given in Table~\ref{tab:constants}.
\begin{turnpage}
   \begin{table*}
      \centering
      \begin{tabular}{ldddccccccc}
         \hline\hline
         method & \multicolumn{1}{c}{exp.~(this work)}
         & \multicolumn{1}{c}{exp.~(ref.~\cite{Reese:JACS126:11387})}
         & \multicolumn{1}{c}{B3PW91} & \multicolumn{1}{c}{B3PW91}
         & \multicolumn{1}{c}{B3LYP} & \multicolumn{1}{c}{B3LYP}
         & \multicolumn{1}{c}{MP2} & \multicolumn{1}{c}{MP2}
         & \multicolumn{1}{c}{CASSCF(8,10)\footnotemark[1]} \\
         basis set & &
         & \multicolumn{1}{c}{6-31G$^*$} & \multicolumn{1}{c}{aug-cc-pVTZ}
         & \multicolumn{1}{c}{6-31G$^*$} & \multicolumn{1}{c}{aug-cc-pVTZ}
         & \multicolumn{1}{c}{6-31G$^*$} & \multicolumn{1}{c}{aug-cc-pVTZ}
         & \multicolumn{1}{c}{6-31G$^*$} \\
         \hline
         cis-3-aminophenol \\%         exp               Pitt           B3PW91         B3LYP         MP2          CASSCF
         \hline
         $A$ / MHz                   & 3734.930\,(14)  & 3734.4\,(7)  & 3741 & 3766 & 3728 & 3755 & 3732 & 3748 & 3748 \\
         $B$ / MHz                   & 1823.2095\,(64) & 1823.1\,(1)  & 1825 & 1837 & 1817 & 1829 & 1815 & 1825 & 1830 \\
         $C$ / MHz                   & 1226.493\,(11)  & 1226.6\,(1)  & 1228 & 1236 & 1223 & 1231 & 1223 & 1229 & 1231 \\
         $\chi_{aa}$ / MHz           & 2.2776\,(34)    &             \\
         $\chi_{bb}-\chi_{cc}$ / MHz & 6.179\,(14)     &             \\
         $\chi_{bb}$ / MHz           & 1.951           &             \\
         $\chi_{cc}$ / MHz           &-4.228           &             \\
         $P^{0}_{cc}$ (u\AA$^2$)     & 0.2262\,(19)    & 0.26(2)     \\
         $\sigma$ / MHz              & 0.002931        & 1.77        \\
         number of lines             & 12              & 114         \\[1ex]
         trans-3-aminophenol \\
         \hline
         $A$ / MHz                   & 3730.1676\,(14) & 3729.5\,(4)  & 3737 & 3766 & 3723 & 3752 & 3715 & 3737 & 3748 \\
         $B$ / MHz                   & 1828.25774\,(58)& 1828.2\,(1)  & 1829 & 1840 & 1821 & 1833 & 1823 & 1832 & 1833 \\
         $C$ / MHz                   & 1228.1948\,(10) & 1228.2\,(1)  & 1229 & 1237 & 1224 & 1232 & 1225 & 1231 & 1232 \\
         $\chi_{aa}$ / MHz           & 2.2666\,(15)    &             \\
         $\chi_{bb}-\chi_{cc}$ / MHz & 6.2405\,(48)    &             \\
         $\chi_{bb}$ / MHz           & 1.987           &             \\
         $\chi_{cc}$ / MHz           &-4.254           &             \\
         $P^{0}_{cc}$ (u\AA$^2$)     & 0.21484\,(18)   & 0.23(2)     \\
         $\sigma$ / MHz              & 0.001647        & 1.61        \\
         number of lines             & 16              & 155         \\
         \hline\hline
      \end{tabular}
      \footnotetext[1]{Private communication with Markus Gerhards (2007); the values are determined
         from the calculations described in reference~\cite{Unterberg:CP304:237}.}
      \caption{Rotational constant, $^{14}$N quadrupole coupling constants, planar moments
         of inertia $P^{0}_{cc}$, overall standard-deviations $\sigma$ of the fit, and the number of
         lines included in the fit for cis- and trans-3-aminophenol.}
      \label{tab:constants}
   \end{table*}
\end{turnpage}
For comparison also the values determined from the high-resolution electronic excitation
spectrum~\cite{Reese:JACS126:11387} and the \emph{ab initio} calculated rotational constants are
shown. The agreement with the previous experimental values is good. The remaining differences might
be attributed to the fact that we determined the rotational constants only from transition between
the very lowest rotational states, where centrifugal distortion effects are negligible. The previous
electronic excitation spectroscopy experiment determined an average value over many more
transitions, including transitions between much higher rotational states, where centrifugal
distortion for such molecules is appreciable. Since, in that work, a rigid-rotor Hamiltonian was
still used to fit the data, the thus obtained rotational constants are effective rotational
constants including these centrifugal distortion effects in an averaged way.

The \emph{ab initio} rotational constants agree reasonably well with the experimental results but
show a considerable spread for the different methods and basis sets. It is obvious, that it would
not be possible to determine the absolute configuration (cis or trans) of the two conformers from
comparisons of experimental and theoretical rotational constants, because the experimental constants
are very similar and the potential errors of the calculated rotational constants are too large for
such an assignment. The planar moments of inertia\footnote{The planar moments $P_{gg}$ of a molecule
   calculate as $2 \cdot P^v_{gg} = I^v_{g'}+I^v_{g''}-I^v_g$. With $g = c, g',g'' = a,b$, and $v =
   0$ we get the planar moment $P_{cc}$ with respect to the $ab$-plane of the principal axes of
   inertia of the oblate rotors in their vibrational ground state:
   $2\cdot{}P^0_{cc}=I^0_a+I^0_b-I^0_c$} of the vibrational ground state of
$P^0_{cc}=0.2262$~u\AA$^2$ for c3AP and $P^0_{cc}=0.21484$~u\AA$^2$ for t3AP are comparable to the
one of aniline ($P^0_{cc}=0.2557$~u\AA$^2$~\cite{Lister:JMolStruct23:253}). Phenol has a
considerably smaller planar moment of $P^0_{cc}=0.01524$~u\AA$^2$~\cite{Larsen:JMolStruct51:175}.
These values indicate that the NH$_2$ configurations of c3AP and t3AP are similar to the one of
aniline and no additional OH out-of-plane contributions compared to phenol are present.

The quadrupole coupling constants are also similar for the two conformers and compared to
aniline~\cite{Kleiboemer:ZNatA43:561}, showing a similar chemical environment of the nitrogen
nucleus in all three molecules. They would not allow to determine the absolute conformation, as was
done for amino acids before~\cite{Lesarri:ACIE43:605}. We have also calculated the barriers to
inversion of the amine-group at the MP2/cc-VTZ level of theory as the energy difference between the
minimum energy for a planar optimized geometry ($C_s$-symmetry) and the absolute minimum on the PES.
The barrier heights obtained in that way are 604~cm$^{-1}$ and 567~cm$^{-1}$ for c3AP and t3AP,
respectively. For comparison we calculated the barrier for aniline at the same level of theory,
which yields a value of 585~cm$^{-1}$ in fair agreement with the experimental values of
525~cm$^{-1}$~\cite{Larsen:CPL43:584, Kydd:CPL49:539} or
540~cm$^{-1}$~\cite{Kleiboemer:ZNatA43:561}, considering that the calculations neglect zero-point
vibrational effects. Overall it is clear that the chemical environment of the amino-group is quite
similar for both conformers of 3AP and for aniline.

% barrier for 3-fluoro-aniline: theory: 518~cm$^{-1}$, experiment: 

\begin{turnpage}
   \begin{table*}
      \centering
      \begin{tabular}{ldddcccccccc}
         \hline\hline
         method & \multicolumn{1}{c}{exp.~(this work)}
         & \multicolumn{1}{c}{exp.~(ref.~\cite{Reese:JACS126:11387})}
         & \multicolumn{1}{c}{vector}
         & \multicolumn{1}{c}{B3PW91} & \multicolumn{1}{c}{B3PW91}
         & \multicolumn{1}{c}{B3LYP} & \multicolumn{1}{c}{B3LYP}
         & \multicolumn{1}{c}{MP2} & \multicolumn{1}{c}{MP2}
         & \multicolumn{1}{c}{CASSCF(8,10)\footnotemark[1]} & \multicolumn{1}{c}{CASPT2\footnotemark[1]} \\
         basis set & &
         & \multicolumn{1}{c}{model}
         & \multicolumn{1}{c}{6-31G$^*$} & \multicolumn{1}{c}{aug-cc-pVTZ}
         & \multicolumn{1}{c}{6-31G$^*$} & \multicolumn{1}{c}{aug-cc-pVTZ}
         & \multicolumn{1}{c}{6-31G$^*$} & \multicolumn{1}{c}{aug-cc-pVTZ}
         & \multicolumn{1}{c}{6-31G$^*$} & \multicolumn{1}{c}{6-31G$^*$} \\
         \hline
         cis-3-aminophenol \\%         exp               Pitt          vec    B3PW91        B3LYP         MP2           CASSCF CASPT2
         \hline
         $\mu_a$ / D                 & 1.7718\,(65)    & 1.77\,(6)   & 1.70 & 1.85 & 1.85 & 1.81 & 1.86 & 1.72 & 1.68 & 1.68 & 1.73 \\
         $\mu_b$ / D                 & 1.517\,(10)    & 1.5\,(2)    & 1.55 & 1.78 & 1.69 & 1.72 & 1.65 & 1.54 & 1.48 & 1.32 & 1.29 \\
         $|\mu|$ / D                 & 2.333\,(11)    & 2.3\,(2)    & 2.30 & 2.57 & 2.51 & 2.50 & 2.49 & 2.31 & 2.14 & 2.14 & 2.16 \\
         $\sigma$ / MHz              & 0.001294 \\
         number of lines             & 10\\[1ex]
         trans-3-aminophenol \\
         \hline
         $\mu_a$ / D                 & 0.5563\,(17)    & 0.57\,(1)   & 0.49 & 0.44 & 0.56 & 0.42 & 0.57 & 0.23 & 0.37 & 0.16 & 0.20 \\
         $\mu_b$ / D                 & 0.5375\,(32)    & 0.5\,(1)    & 0.56 & 0.51 & 0.43 & 0.54 & 0.48 & 0.81 & 0.63 & 1.05 & 1.09 \\
         $|\mu|$ / D                 & 0.7735\,(34)    & 0.7\,(1)    & 0.74 & 0.67 & 0.71 & 0.68 & 0.75 & 0.84 & 0.73 & 1.06 & 1.11 \\
         $\sigma$ / MHz              & 0.000890 \\
         number of lines             & 19\\[1ex]
         \hline\hline
      \end{tabular}
      \footnotetext[1]{Private communication with Markus Gerhards (2007); the values are determined from
         the calculations described in reference \cite{Unterberg:CP304:237}. The CASPT2 dipole moments
         are calculated at the CASSCF-optimized geometry.}
      \caption{Dipole moment components $\mu_a$ and $\mu_b$, overall dipole moments $|\mu|$ of cis- and
         trans-3-aminophenol, standard deviations $\sigma$ of the fits, and the number of
         measurements at different field strengths included in the fit. See text for details.}
      \label{tab:dipole}
   \end{table*}
\end{turnpage}
From the Stark shift measurements we precisely determined the $\mu_a$ and $\mu_b$ dipole moment
components of the two conformers by fitting the transition shifts with
QStark~\cite{Kisiel:CPL325:523, Kisiel:JPC104:6970}. In the fitting procedure we set up and directly
diagonalized the $M$-matrices,\footnote{In all calculations the matrix dimensions in $J$ and $F$ exceeded their maximum
   values in the experimental data by 10.} yielding an accurate description of the Stark effect for
all field strengths. The experimental dipole moments are given in Table~\ref{tab:dipole} together
with \emph{ab initio} values. We also calculated the dipole moments of c3AP and t3AP from the dipole
moments of aniline \cite{Lister:JMolStruct23:253} and phenol \cite{Larsen:JMolStruct51:175} using
simple vector addition of the individual dipole moments in the aminophenol principal axes system.
These values are in excellent agreement with the experimentally observed values, confirming the
dipole additivity for the ground states of 3AP \cite{Borst:CPL350:485, Reese:JACS126:11387}. Using
either the dipole moments predicted by the vector model or by \emph{ab initio} calculations, the
absolute conformation of the molecules can easily be obtained. The assignment agrees with the
previous Stark effect study by Reese et al.~\cite{Reese:JACS126:11387}. This procedure of
distinguishing conformers based on their dipole moment orientation is quite comparable to the use of
vibrational transition moment angles \cite{Dong:Science298:1227}. In that work the angle between
permanent dipole moment and vibrational transition dipole moments was used to distinguish different
conformers. Here we use the angle between the permanent dipole moment and the principal axes for the
same purpose. Overall, it is a very helpful tool to employ experimental dipole moments to determine
the absolute conformational structure of modular molecules.

The quality of the \emph{ab initio} dipole moments is quite unsatisfactory. The calculated dipole
moments for the more polar conformer c3AP are typically within 10~\%. For the less polar conformer
t3AP the calculated values differ by considerably more than 10~\% from the experimental values
for all levels of theory. However, what is most discouraging for the calculation of dipole moments
of large and modular molecules is the wide distribution of their calculated orientations, as can be
seen from the individual components along the principal axes. Moreover, increasing the level of
theory and using the larger basis set often even gives worse results. For the CASSCF calculations
systematic problems in the descriptions of hydroxybenzenes due to the neglect of
$\sigma$-correlation have already been described~\cite{Unterberg:CP304:237}. Density-functional
theory (esp.\ the B3LYP functional) using a large basis set (aug-cc-pVTZ) gives the best overall
results for the two conformers of 3AP, but it has to be more carefully examined for a wider class of
molecules whether this is serendipitous or a consistent quality of this method. Overall it must be
concluded, that the \emph{ab initio} methods used here, which are applied routinely in the
calculation of structures and vibrational frequencies of molecules of similar size as 3AP, perform
quite poorly for the calculation of electric dipole moments of these molecules. Therefore, one has
to be careful when using such \emph{ab initio} results for the calculation of Stark shifts of large
molecules, as the dipole moment components directly enter the Hamiltonian matrix calculation
\cite{Gordy:MWMolSpec}.

The dipole moments also yield important information on the intermolecular interactions of molecules.
However, in a recent study of the 3AP water cluster, only a single dimer structure could be
observed, an OH--OH$_2$ bound hydrogen structure between the t3AP conformer and water. No c3AP-water
cluster was found \cite{Gerhards:JCP123:074320}. Considering the much larger dipole moment of c3AP
this indicates that the local OH--OH$_2$ interaction and steric effects, which are favorable for
t3AP-water, are dominating over the overall electrostatic interaction. For bulk solutions, on the
other hand, we would expect the more polar c3AP conformer to have a stronger interaction with the
solvent than the less polar t3AP.

\section{Conclusions}
\label{sec:conclusions}

We measured several hyperfine-resolved rotational transitions of the conformers of 3AP for the
lowest $J$ values using FTMW spectroscopy. From an rigid rotor analysis we could precisely determine
the rotational constants and $^{14}$N nuclear quadrupole coupling constants. These nuclear
quadrupole coupling constants and the calculated barriers to inversion are comparable to the values
of aniline, confirming a similar electronic configuration around the nitrogen nucleus. In order to
assign the absolute conformation of the individual species, the rotational constants and the
quadrupole coupling constants could not be used, as they are too similar for the two conformers.
However, we have very precisely determined the dipole moments of both conformers using the Stark
shifts of several of these lines. These dipole moments can be rationalized in terms of simple vector
addition of the dipole moments of aniline and phenol, and they allow to unambiguously assign the
absolute conformation, even without the need for \emph{ab initio} calculations.

The precisely known rotational constants and dipole moment components allow us to calculate the
Stark effect of the individual rotational states with good accuracy even for very high electric
fields (i.\,e.~200~kV/cm), as applied in Alternate Gradient decelerators \cite{Bethlem:JPB39:R263}
or similar focusing devices. The Stark shifts of the two conformers are quite different, with
effective dipole moments of 0.0365~$\text{cm}^{-1}/(\text{kV\,cm}^{-1})$ for c3AP and
0.0113~$\text{cm}^{-1}/(\text{kV\,cm}^{-1})$ for t3AP.\footnote{A dipole moment of 1~D corresponds
   to 0.0168~$\text{cm}^{-1}/(\text{kV\,cm}^{-1})$.} Therefore one can envision to spatially
separate the conformers of aminophenol or similar modular (bio-)molecules by exploiting these Stark
effect differences in an $m/\mu$-selector, similar to the $m/q$-selection in a mass spectrometer for
charged particles.

\begin{acknowledgments}
   We would like to thank Gerard Meijer for his continuous support of this work and for helpful
   discussions. We thank Boris Sartakov for helpful discussions on magnetic hyperfine interactions.
   Financial support from the \emph{Land Niedersachsen} and the \emph{Deutsche
      Forschungsgemeinschaft}, also within the priority program 1116 ``Interactions in ultra-cold
   atomic and molecular gases'', is acknowledged.
\end{acknowledgments}

\appendix
\section{Calibration of the Stark effect measurements}
\label{sec:calibration}

The most critical factor in the determination of electric dipole moments of molecules by Stark
effects measurements are accurate values of the applied electric fields. Often sufficiently
homogeneous electric fields are generated by applying a voltage difference using two plane-parallel
electrodes~\cite{Gordy:MWMolSpec}. For the large volumes sampled in FTMW spectrometers this is
difficult and, therefore, in the Hannover COBRA-FTMW experiment a different approach is taken:
Within the CAESAR setup, the two concave mirrors of the spectrometer resonator are used as the
primary capacitor electrodes and electrode rings between them are used to create a homogeneous field
over a very large fraction of the resonator volume~\cite{Schnell:RSI75:2111}. Therefore, all
detected molecules experience practically the same field strength, even for the large modes at
frequencies below 3~GHz.

In all such experiments the field strength must be calibrated, which is often performed in turn by
Stark effect measurements for isotopologues of OCS, for which the Stark effect is known with high
precision~\cite{Gordy:MWMolSpec}. Such measurements yield an averaged effective parallel-plate
separation $d$, which can then be used for the calculation of the electric field strength applied
during other measurements performed in the same setup. In our setup, however, the cavity mirrors
serve as main electrodes. For different measurement frequencies these mirrors need to be moved to
fulfill the resonance conditions of the cavity. Since the effective $d$ has a slight dependence on
the mirror separation, we apply a linear correction for this effect to all our measurements.

We first determine effective $d$s from Stark effect measurements on OC$^{36}$S, assuming a dipole
moment of 0.71519\,(3)~D~\cite{Reinartz:CPL24:346} and using a linear calibration curve to account
for systematic errors due to a discrepancy between the displayed and the actual voltage applied to
the electrodes using an FuG HCD-20000 high-voltage power supply: $U/V = 3.86\thinspace{}(70) + 1.00203\thinspace{}(15) \cdot
U_{disp}/V$. The effective parallel-plate separation $d$ is determined at two very different
relative mirror positions, namely $z_1=7.12$~mm and $z_2=46.13$~mm. These positions are specified
relative to the zero-position of the moving cavity mirror and correspond to two quite extreme
translational positions of the mirror. For these $z$-positions we determine the effective plate
separation to $d_1=0.58343\thinspace{}(31)$~m and $d_2=0.60099\thinspace{}(55)$~m, respectively.

For all Stark effect measurements on 3AP we then note the actual $z$-position and determine the
effective $d$ for that measurement by linear interpolation between the two calibration-measurements.
The estimated error in the $z$-position of 0.01~mm for all measurements is much smaller than the
error of $d$ and has thus been neglected. In order to account for the contribution of systematic
errors of the electric field strength to the total error of the dipole moments, fits were performed
at the extreme values of the electric field as derived from the calibration of the voltage and of
the effective plate separation (\emph{vide supra}). These fits provide upper and lower bounds for
the dipole moments of 3-aminophenol due to systematic errors, which are included in the error
estimates given in Table~\ref{tab:dipole}.

\bibliographystyle{apsrev-nourl}
\bibliography{string,mp}

\end{document}